\documentclass[11pt,a4paper]{article}

\usepackage[utf8]{inputenc}
\usepackage[T1]{fontenc}
\usepackage{amsmath,amssymb}
\usepackage{graphicx}
\usepackage{hyperref}
\usepackage{booktabs}
\usepackage{array}
\usepackage{geometry}
\title{Soppia: A Structured Prompting Framework for the Proportional Assessment of Non-Pecuniary Damages in Personal Injury Cases}

\author{
    Jorge Alberto Araujo\\
    Tribunal Regional do Trabalho da 4ª Região (TRT4), Porto Alegre, Brazil\\
    \texttt{jaaraujo@trt4.jus.br}
}

\date{}

\begin{document}

\maketitle

\begin{abstract}
Applying complex legal rules characterized by multiple, heterogeneously weighted criteria presents a fundamental challenge in judicial decision-making, often hindering the consistent realization of legislative intent. This challenge is acutely evident in the quantification of non-pecuniary damages in personal injury cases. This paper introduces \textbf{Soppia}---a structured prompting framework designed to assist legal professionals in navigating this complexity. By leveraging advanced AI, the system ensures a comprehensive, balanced analysis of all stipulated criteria, faithfully fulfilling the legislator’s intent that compensation be determined through a holistic assessment of the specific circumstances of each case. Using the 12 criteria for non-pecuniary damages established in the Brazilian \textbf{CLT}\footnote{Consolidação das Leis do Trabalho (Consolidation of Labor Laws)---Brazil's primary legislative body governing labor relations.} (Art. 223-G) as a case study, we demonstrate how Soppia (\textbf{S}ystem for \textbf{O}rdered \textbf{P}roportional and \textbf{P}ondered \textbf{I}ntelligent \textbf{A}ssessment) operationalizes nuanced legal commands into a practical, replicable, and transparent methodology. This approach not only enhances consistency and predictability but also establishes a versatile tool for all legal practitioners. We argue that the methodology is highly adaptable across multi-criteria legal contexts, thereby promoting a more uniform and equitable application of justice. Soppia thus bridges normative interpretation and computational reasoning, paving the way for explainable, auditable legal AI frameworks.

\bigskip

\end{abstract}

\section{Introduction}
The assessment of compensation for non-pecuniary damages, such as pain, suffering, and emotional distress, is one of the most complex tasks in the administration of justice. In the absence of a clear monetary equivalent, these damages are often quantified based on the subjective discretion of judges, leading to significant variability in outcomes for similar cases. This inconsistency, or “noise,” as described by Kahneman, Sibony, and Sunstein~\cite{kahneman2021noise}, undermines the predictability of judicial decisions and can erode public trust in the judiciary. The challenge is universal, affecting tort law, consumer protection, and human rights litigation alike.

The demand for transparency and interpretability in AI-based judicial support tools has been repeatedly emphasized in recent scholarship.
Deeks~\cite{deeks2019} argues that courts increasingly require AI systems to provide explainable outputs consistent with procedural fairness,
while Richmond et al.~\cite{richmond2024} highlight that explainability in legal AI is not only a technical concern but an epistemic one,
deeply connected to how judges justify decisions. These perspectives reinforce the need for frameworks such as Soppia, which make explicit the logical structure underlying judicial reasoning.

In recent years, the emergence of Large Language Models (LLMs) and other forms of Artificial Intelligence (AI) has opened new possibilities for supporting complex decision-making processes. However, the use of AI in law, especially in high-stakes judicial contexts, raises critical questions about transparency, accountability, and the preservation of human oversight. The concept of ``black-box'' models, whose internal logic is opaque, is particularly problematic in a field that demands reasoned and justifiable outcomes~\cite{rudin2019stop}.

This paper addresses these challenges by proposing \textit{Soppia}—a structured prompting framework for the proportional assessment of non-pecuniary damages. Our methodology, embodied in this codename (\textbf{S}ystem for \textbf{O}rdered \textbf{P}roportional and \textbf{P}ondered \textbf{I}ntelligent \textbf{A}ssessment), is not intended to replace human judges but to augment their decision-making process with a transparent, consistent, and auditable tool~\cite{cardoso2024,unesco2024,agarwal2022,cnj2025}. The name `Soppia' evokes simplicity, approachability, and reliability---essential values for a tool designed to support complex judicial decisions.

By structuring the analysis around a predefined set of weighted criteria, Soppia guides both human and AI reasoning, ensuring that all relevant factors are systematically considered. To illustrate its practical application, we use the comprehensive set of 12 criteria for assessing non-pecuniary damages in workplace accidents, as stipulated by Brazilian law. This serves as a robust case study for constructing a detailed and replicable prompt that can be adapted for use in any jurisdiction.

We demonstrate how this tool can be used not only by judges to draft more consistent and well-founded rulings but also by lawyers to build stronger arguments, by companies for risk management, and by academics for legal analysis. Ultimately, this work aims to contribute to the development of explainable AI (XAI) in the legal domain~\cite{richmond2024}, fostering a more equitable and predictable system of justice.

\section{Theoretical Foundations}

Our framework is built upon three fundamental pillars: the theory of decision-making, the principles of explainable artificial intelligence, and the practice of prompt engineering.

\subsection{Noise and Bias in Human Judgment}

The seminal work of Daniel Kahneman, Olivier Sibony, and Cass R. Sunstein, \textit{Noise: A Flaw in Human Judgment}, provides the primary theoretical impetus for this research~\cite{kahneman2021noise}. They define noise as the ``undesirable variability in judgments of the same problem.'' In the legal context, this manifests when different judges award widely varying compensation amounts for similar injuries and circumstances. Our framework acts as a form of "decision hygiene," a strategy to reduce noise by structuring the decision-making process. By breaking down the complex decision of quantifying damages into a series of smaller, more manageable assessments based on explicit criteria, the framework helps ensure that all decision-makers are evaluating the same factors in a comparable manner.

\subsection{Explainable AI (XAI) and Interpretable Models}

The use of AI in law demands a departure from opaque, "black-box" models. Cynthia Rudin argues forcefully against merely trying to "explain" black boxes, advocating instead for the creation of inherently interpretable models~\cite{rudin2019stop}. This conception of explainable reasoning aligns with Bench-Capon and Sartor’s classic model of legal reasoning,
which integrates factual analysis with normative theories and underlying values~\cite{benchcapon2001}.
An interpretable model is one whose decision-making process is understandable to a human~\cite{richmond2024}. Our framework aligns with this principle. It does not rely on a complex, unexplainable algorithm to reach a conclusion. Instead, it uses the LLM as a tool to process information within a transparent, human-designed structure. The final output is not a mysterious number but a detailed, step-by-step analysis that is fully auditable and comprehensible. This approach meets the judicial demand for explainable AI, as it allows judges and litigants to understand precisely how a conclusion was reached~\cite{deeks2019}.

This conception of explainable reasoning is consistent with Bench-Capon and Sartor’s model of legal reasoning,
which integrates factual analysis with normative theories and values~\cite{benchcapon2001}.

\subsection{Prompt Engineering for Legal Applications}

A prompt is a set of instructions given to an LLM to guide its output. Effective prompt engineering is crucial for harnessing the power of these models for specialized tasks. Our framework is, in essence, a sophisticated and highly structured prompt. It translates a complex legal task---the dosimetry of damages---into a clear set of instructions that the AI can follow. This includes defining the criteria for analysis, specifying the scoring scales, providing the weighting system, and describing the desired output format. In doing so, the prompt ensures that the powerful analytical capabilities of the AI are channeled in a manner consistent with legal principles and judicial requirements.

\section{Methodology: The Structured Prompt Framework}

The framework is designed to be both systematic and flexible. It consists of four main components: a defined set of criteria, a calibrated scoring system, a weighting mechanism, and a classification structure.

Following the interpretability framework proposed by Doshi-Velez and Kim~\cite{doshi2017},
Soppia adopts a structured prompting approach that prioritizes transparency and reproducibility over opaque, performance-oriented models.

\subsection{Identification and Definition of Criteria}

The first step is to identify the legally relevant criteria for assessing the damage. These criteria should be derived from statutes, case law, or legal doctrine. For our case study, we use the 12 criteria from Article 223-G of the Brazilian Consolidation of Labor Laws, which provides a comprehensive list of factors to be considered in cases of non-pecuniary damages. These are presented in Table~\ref{tab:criteria}.

\begin{table}[ht]
\centering
\caption{The 12 Criteria for Non-Pecuniary Damages Assessment (Art. 223-G, CLT)}
\label{tab:criteria}
\small
\begin{tabular}{@{}clp{7cm}@{}}
\toprule
\textbf{\#} & \textbf{Criterion} & \textbf{Description} \\ 
\midrule
I & Nature of the protected legal interest & The importance of the violated right (e.g., life, health, dignity). \\
II & Intensity of suffering or humiliation & The degree of physical and psychological pain experienced by the victim. \\
III & Possibility of physical or psychological recovery & The likelihood of the victim overcoming the consequences of the damage. \\
IV & Personal and social repercussions of the damage & The impact of the damage on the victim's family, social, and professional life. \\
V & Extent and duration of the effects of the damage & The persistence of the consequences of the damage over time (temporary vs. permanent). \\
VI & Conditions under which the offense occurred & The specific circumstances of the harmful event. \\
VII & Degree of intent or fault & The level of culpability of the offender (negligence, recklessness, or intent). \\
VIII & Spontaneous retraction & Whether the offender apologized or acknowledged the wrongdoing. \\
IX & Effort to mitigate the damage & Actions taken by the offender to assist the victim after the event. \\
X & Express or tacit forgiveness & Whether the victim has forgiven the offender. \\
XI & Economic situation of the parties & The financial capacity of the offender and the needs of the victim. \\
XII & Degree of publicity of the offense & The extent of public exposure of the harmful event. \\
\bottomrule
\end{tabular}
\end{table}

\subsection{Calibrated Scoring System with Dual Logic}

Each criterion is evaluated on a scale of 1 to 5 points. A key innovation of our framework is the use of a dual-logic system to ensure that a higher score consistently reflects greater severity of the damage, thus justifying a higher compensation amount.

\begin{itemize}
    \item \textbf{Direct Logic (8 criteria):} For most criteria, a higher score is assigned to a greater presence of the factor. For example, a greater intensity of suffering (Criterion II) results in a higher score.
    \item \textbf{Inverse Logic (4 criteria):} For mitigating factors, the logic is inverted. A lower presence of the factor results in a higher score. For example, a complete lack of effort to mitigate the damage (Criterion IX) receives a score of 5, as this absence aggravates the overall damage.
\end{itemize}

This dual-logic system is crucial for correctly modeling legal reasoning, in which both aggravating and mitigating circumstances must be weighed.

\subsection{Weighting System}

Not all criteria are of equal importance. The framework introduces a weighting system to reflect the relative relevance of each factor. For example, the possibility of recovery (Criterion III) and the duration of the effects (Criterion V) receive higher weights because they are decisive in distinguishing temporary damages from permanent, life-altering injuries. Conversely, factors that are rarely applicable or have a lesser impact, such as forgiveness (Criterion X), receive a lower weight. The complete weighting system is presented in Table~\ref{tab:weights}.

\begin{table}[ht]
\centering
\caption{Weighting System for the 12 Criteria}
\label{tab:weights}
\small
\begin{tabular}{@{}clp{8cm}@{}}
\toprule
\textbf{Criterion} & \textbf{Weight} & \textbf{Justification} \\ 
\midrule
III & 2.5× & Decisive: Differentiates temporary damages from permanent and disabling conditions, directly impacting the projection of future suffering. \\
V & 2.0× & Very Important: The persistence of the consequences over time is a central factor in quantifying the intensity and perpetuity of the harm. \\
I & 1.5× & Fundamental: The violation of rights such as psychological integrity or honor is inherently more severe than that of other legal interests. \\
II & 1.5× & Very Important: Represents the core of the subjective experience of non-pecuniary damage, assessing the degree of pain and anguish actually endured. \\
VII & 1.2× & Important: The reprehensibility of the offender's conduct (negligence, recklessness, or intent) is a pillar of civil liability. \\
IV & 1.0× & Standard Importance: Assesses the concrete impact of the damage on the victim's personal, social, and professional life. \\
VI & 1.0× & Standard Importance: The context in which the offense occurred (e.g., public, violent) can significantly aggravate the humiliation. \\
XI & 1.0× & Standard Importance: Seeks to balance the reparation so that it is compensatory for the victim without becoming confiscatory for the offender. \\
X & 1.0× & Key Relational Factor: The absence of forgiveness suggests that the damage has not been overcome, aggravating the offense. \\
IX & 0.8× & Relevant Mitigating Factor: Concrete actions by the offender to assist the victim after the event demonstrate responsibility and can reduce the intensity of suffering. \\
VIII & 0.6× & Minor Mitigating Factor: An apology has a limited impact if it is not accompanied by concrete actions or is rejected by the victim. \\
XII & 0.5× & Generally Low Impact: In cases of pure non-pecuniary damage, publicity has less weight, unless it is a central and proven aggravating element. \\
\bottomrule
\end{tabular}
\end{table}

The final score for a case is the weighted sum of the scores for each criterion.

\subsection{Classification and Final Adjustment}

The total weighted score is then used to classify the overall severity of the damage into one of four categories: Mild, Medium, Severe, or Very Severe. Each category corresponds to a predefined compensation range, often expressed as a multiple of a relevant baseline (e.g., the victim's salary or the minimum wage).

Finally, a fine-tuning mechanism allows for modulation within the selected range based on the score's position (lower, middle, or upper third of the range), providing a more granular and proportional final recommendation.

All prompts and implementation materials used in the Soppia framework are openly available for consultation and replication at:
\url{https://github.com/jaa41/soppia-framework}.

\section{The Framework as a Replicable Prompt}

The methodology described above can be encapsulated in a detailed prompt for an LLM. This prompt instructs the AI to perform the analysis step-by-step, ensuring a transparent and well-documented output. The complete and continuously updated prompt, along with usage examples and templates, is publicly available in this project's GitHub repository. For immediate illustration, a simplified version is provided in Appendix A, but its core structure guides the AI to:

\begin{enumerate}
    \item \textbf{Receive Specific Case Information:} The user provides the facts of the case, including medical reports, witness testimonies, and other evidence.
    \item \textbf{Analyze Each of the 12 Criteria:} For each criterion, the AI must justify its assessment based on the evidence provided and assign a score from 1 to 5, explicitly stating the logic (direct or inverse) being applied.
    \item \textbf{Calculate the Weighted Score:} The AI multiplies each score by its corresponding weight and sums the results.
    \item \textbf{Classify the Damage:} Based on the total weighted score, the AI classifies the damage into one of the four severity levels.
    \item \textbf{Propose a Compensation Range:} The AI suggests a compensation range based on the classification.
    \item \textbf{Generate a Justification Report:} The AI produces a complete report detailing the entire analysis, which can be used as the basis for a legal argument or judicial decision.
\end{enumerate}

This dual availability---a static resource in the appendix for verification and a dynamic resource on GitHub for practical application---aims to maximize the transparency, replicability, and ongoing utility of the Soppia framework.

\section{Applications and Reproducibility}

While our case study is based on Brazilian law, the framework's true strength lies in its adaptability. Judges can use the generated report to structure and justify their decisions. Lawyers can employ the framework to assess the strength of a case, manage client expectations, and formulate settlement strategies. Companies can use it as a proactive risk management tool, understanding the potential liability associated with different types of incidents.

The framework can be readily adapted to other contexts. In consumer law, the criteria could be modified to include factors such as consumer vulnerability or the deceptive nature of a commercial practice. In environmental law, the criteria could include the extent of ecological damage and the reversibility of the harm. Legal professionals in any jurisdiction can adapt the framework by replacing the Brazilian criteria with the relevant statutory or common law factors in their own legal system and calibrating the weights to reflect local legal culture and precedents.

\section{Conclusion}

The Soppia Framework stands as a practical and methodological contribution to the dosimetry of non-pecuniary damages, offering a replicable path toward more consistent, transparent, and well-founded judicial decisions. By combining the structured logic of legally validated criteria with the analytical processing power of AI, the system demonstrates that it is possible to significantly reduce judicial noise without sacrificing human oversight, explainability, or the necessary individualization of analysis.

Ensuring proportionality and fairness in AI-assisted judgments is essential to maintain judicial integrity.
Bharati~\cite{bharati2025} emphasizes that bias mitigation and fairness auditing should be integral components of any legal AI system.

As a magistrate who uses this tool in my own judicial activity, I attest that its primary value is not as a “magic formula” but as an instrument for organizing the decision-making process. The proposed weighting for the criteria in Article 223-G of the CLT, calibrated from practical experience and direct contact with case law, serves as a robust but intentionally debatable and adaptable starting point. Lawyers can—and should—argue for the revision of specific weights in their pleadings, just as other judges can recalibrate the model in light of their own convictions and local context.

However, inherent limitations are acknowledged. The effectiveness of Soppia remains tied to the quality of the information provided to the prompt, and its structure, optimized for Brazilian labor law, will require adjustments for transposition to other jurisdictions or branches of law.

As emphasized by Deeks~\cite{deeks2019}, the judiciary increasingly demands explainable systems
capable of justifying their reasoning.
Soppia directly addresses this demand by providing a transparent, auditable, and normatively grounded framework for proportional legal assessment.

The true legacy of Soppia, therefore, lies not in providing a definitive answer but in institutionalizing a method. It translates a complex and subjective judgment into a structured, auditable, and adversarial analytical process. As AI irreversibly integrates into the legal profession, frameworks like this will be essential to ensure that technological advancement always serves the fundamental principles of justice: fairness, proportionality, and rational justification.

\newpage
\appendix

\section{Simplified Version of the Soppia Prompt}

\noindent\textbf{ROLE:} You are the Soppia legal assistant. Your task is to perform a structured analysis of non-pecuniary damages according to Art. 223-G of the CLT.

\noindent\textbf{CONTEXT:} The user will provide the facts of a workplace accident or occupational disease case.

\noindent\textbf{INSTRUCTIONS:}
\begin{enumerate}
    \item Analyze each criterion of Art. 223-G based on the evidence provided.
    \item Assign a score from 1 to 5 for each criterion, applying direct or inverse logic as specified.
    \item Calculate the total weighted score using the established weights.
    \item Classify the severity of the damage (Mild, Medium, Severe, Very Severe).
    \item Suggest a compensation range based on the classification.
    \item Generate a justifying report detailing the entire analysis.
\end{enumerate}

\noindent\textbf{CRITERIA AND WEIGHTS (SUMMARIZED):}

\begin{table}[ht]
\centering
\small
\begin{tabular}{@{}lcc@{}}
\toprule
\textbf{Criterion} & \textbf{Weight} & \textbf{Logic} \\ 
\midrule
III - Possibility of recovery & 2.5× & Inverse \\
V - Extent and duration of effects & 2.0× & Direct \\
I - Nature of the legal interest & 1.5× & Direct \\
II - Intensity of suffering & 1.5× & Direct \\
VII - Degree of intent or fault & 1.2× & Direct \\
IV, VI, XI - Other criteria & 1.0× & Direct \\
X - Forgiveness & 1.0× & Inverse \\
IX - Effort to mitigate & 0.8× & Inverse \\
VIII - Spontaneous retraction & 0.6× & Inverse \\
XII - Publicity of the offense & 0.5× & Direct \\
\bottomrule
\end{tabular}
\end{table}

\noindent\textbf{CLASSIFICATION:}
\begin{itemize}
    \item 15--32 points: MILD Offense (up to 3× salary)
    \item 33--50 points: MEDIUM Offense (up to 5× salary)
    \item 51--68 points: SEVERE Offense (up to 20× salary)
    \item 69+ points: VERY SEVERE Offense (up to 50× salary)
\end{itemize}

\section*{Structured Output Format}

\begin{enumerate}
    \item \textbf{CASE SUMMARY} \\
    Brief description of the facts.

    \item \textbf{CRITERIA ANALYSIS} \\
    For each criterion: analysis, score, and justification.

    \item \textbf{FINAL CALCULATION}
    \begin{itemize}
        \item Total weighted score: [X] points
        \item Classification: [Mild/Medium/Severe/Very Severe]
        \item Suggested compensation: [Range based on salary]
    \end{itemize}

    \item \textbf{CONCLUSION} \\
    Summary of analysis and final recommendation.
\end{enumerate}

\end{document}